\documentclass[copyright]{eptcs}
\providecommand{\event}{T. Neary, D. Woods, A.K. Seda and N. Murphy (Eds.):\\ The Complexity of Simple Programs 2008.}
\providecommand{\volume}{1}
\providecommand{\anno}{2009}
\providecommand{\firstpage}{70}
\usepackage{breakurl}

\usepackage{natbib}

\usepackage[OT1]{fontenc}

\usepackage{xspace}

\usepackage{verbatim}

\usepackage{graphicx}
\graphicspath{{./}}
\DeclareGraphicsExtensions{.jpeg,.jpg,.eps}

\usepackage[center]{subfigure}

\usepackage{h_general}
\usepackage{h_math}
\usepackage{h_ca}
\usepackage{h_agc}
\usepackage{h_graphics}
\usepackage{h_couleur}

\AGCmakeSignal{BouncerLL}{boun$_L$}{DarkPurple}
\AGCmakeSignal[,linewidth=1.2pt]{BorderLOne}{bord$^L_1$}{DarkPurple}
\AGCmakeSignal[,linewidth=1.2pt,linestyle=dashed]{BorderLTwo}{bord$^L_2$}{DarkPurple}
\AGCmakeSignal[,linewidth=1.2pt,linestyle=dotted]{BorderLThree}{bord$^L_3$}{DarkPurple}

\AGCmakeSignal{MetaOne}{$\AGCmetaSignal_1$}{DarkBlue}
\AGCmakeSignal{MetaTwo}{$\AGCmetaSignal_2$}{DarkRed}
\AGCmakeSignal{MetaThree}{$\AGCmetaSignal_3$}{DarkGreen}

\newcommand{\CAstate}{\MathXspace{s}}

\AGCmakeSignal[,linestyle=dashed]{OneL}{one$_L$}{DarkGreen}
\AGCmakeSignal[,linestyle=dashed,dash=1pt 1pt]{ZeroL}{zero$_L$}{Black}

\AGCmakeSignal[,linewidth=1.2pt]{First}{first}{DarkPurple}
\AGCmakeSignal[,linewidth=1.2pt]{Sep}{sep}{Red}
\AGCmakeSignal[,linewidth=1.2pt]{Last}{last}{Black}

\AGCmakeSignal{TrueR}{true$_{R}$}{Blue}
\AGCmakeSignal[,linewidth=1.2pt]{FalseR}{false$_R$}{Red}

\AGCmakeSignal[,linewidth=1.2pt]{GoRR}{go$_{RR}$}{Blue}

\AGCmakeSignal{GoLL}{go$_{LL}$}{Blue}

\AGCmakeSignal[,linestyle=dashed]{One}{one}{DarkGreen}
\AGCmakeSignal[,linestyle=dashed]{OneR}{one$_R$}{DarkGreen}
\AGCmakeSignal[,linestyle=dashed]{OneRR}{one$_{RR}$}{DarkGreen}

\AGCmakeSignal[,linestyle=dashed,dash=1pt 1pt]{Zero}{zero}{Black}
\AGCmakeSignal[,linestyle=dashed,dash=1pt 1pt]{ZeroR}{zero$_R$}{Black}
\AGCmakeSignal[,linestyle=dashed,dash=1pt 1pt]{ZeroRR}{zero$_{RR}$}{Black}

\AGCmakeSignal{Halt}{halt}{DarkRed}
\AGCmakeSignal{HaltR}{halt$_R$}{DarkRed}

\newcommand\T{\rule{0pt}{2.4ex}}
\newcommand\B{\rule[-1.2ex]{0pt}{0pt}}

\title{Small Turing universal signal machines}
\author{J{\'e}r{\^o}me {Durand-Lose}
\institute{Laboratoire d'Informatique Fondamentale d'Orl\'eans,\\ 
  Universit\'e d'Orl\'eans, \\
  B.P.~6759, \\
  F-45067 ORL\'EANS Cedex 2.
}
\email{Jerome.Durand-Lose@univ-orleans.fr}
}

\def\titlerunning{Small Turing universal signal machines}
\def\authorrunning{J. Durand-Lose}

\begin{document}
\maketitle

\begin{abstract}
  This article aims at providing signal machines as small as possible able to perform any computation (in the classical understanding).
  After presenting signal machines, it is shown how to get universal ones from Turing machines, cellular-automata and cyclic tag systems.
  Finally a halting universal signal machine with $13$ meta-signals and $21$ collision rules is presented.
\end{abstract}

%


\section{Introduction}
\label{sec:intro}

Computation and universality have been defined in the 1930's.
In the last five or so decades, it has been unveiled how common they are.
The question about the frontier moved from proving the universality of dynamical systems to the complexity of universal machines.
This is not only an intellectual challenge, but also important to find niches escaping tons of non decidability results or otherwise assert that these niches are too small to be of any interest.

There have already been a lot of investigations on small 
 Turing machines \citep{rogozhin82,rogozhin96,margenstern95,kudlek96tcs,baiocchi01mcu}, 
 register machines \citep{korec96tcs}, and
 cellular automata \citep{ollinger02icalp,cook04:cs}.

Moreover, as (Turing) universality (capability to carry out any Turing/classical computation) has been developed in very limited systems, some ``adaptations'' were made and various notions of universality exist:
\begin{itemize}
\item \emph{polynomial time universality} when polynomial computations (as defined in complexity theory) are still done in polynomial time, as opposed to \emph{exponential time universality} provided by \eg 2-counter automata, and
\item \emph{semi-universality} when the computation must be started on an infinite configuration --for example the whole tape of a Turing machine is filled with some ultimately-periodic infinite world-- or in many cases like cellular automata which naturally work on infinite configurations.
\end{itemize}

\medskip

In the present article, minimal universal machine in the context of \emph{Abstract geometrical computation} (AGC) is investigated.
AGC has been introduced as a continuous counterpart of cellular automata.
This move is inspired by the way dynamics of CA is often designed or analyzed in an Euclidean space. 
In AGC, signals are moving with constant speed in an Euclidean space.
When they meet, they are replaced/rewritten; interacting in a collision based computing way.
A \emph{signal machine} (SM) defines existing kinds of signal, \emph{meta-signals}, and their interactions, \emph{collision rules}.
AGC allows, since this is a graphical model, to understand the way the information is moved around and interacts as shown by the various illustrations.

Abstract geometrical computation uses continuous space and time so that the possibility Zeno effect has to be considered.
Indeed, it can happen and be used to compute beyond Turing computability and to climb the arithmetical hierarchy by using accumulations of collisions \citep{durand.lose09nc}.
But since only Turing computability is addressed here, accumulations are not considered.

The straightforward measure of complexity (or simple/small-ness) of an SM is the number of meta-signals.
If there are \AGCmetaSignalNbr meta-signals defined, then there are at most $2^\AGCmetaSignalNbr-\AGCmetaSignalNbr-1$ possible collision rules (at least two signals are needed in a collision and they must be all different).
Since parallel signals cannot interact, the number of rules could be much lower, but nevertheless exceeding by far the number of meta-signals.
In many constructions, only a small part is defined, the rest being either undefined or \emph{blank} (\ie signals just cross each other). 
So that the number of specially defined rules is also a good complement to the size of an SM.

In \cite{durand.lose05mcu04,durand.lose05cie05}, Turing universality is proven by reduction from 2-counter automata.
This result is not interesting here since the universal SM they provide are exponential time universal and not very small.
(For an SM computation, the \emph{time complexity} is the longest chain of collisions linked by signals.)

\medskip

In the present article, universal SM are generated from Turing machines, cellular automata and cyclic tag systems.
In each case, special care is taken in order to save signals and collision rules.
For Turing machines, only results based on the construction in \cite{durand.lose09nc} are presented; this direct simulation is not detailed since it does not provide the best bound.
Cellular automata (CA) are massively parallel devices where bounds are also very tight and provide a very small semi-universal SM.

Cyclic tag systems (CTS) work by considering a binary word and a circular list of binary words.
At each iteration, the first bit of the word is deleted and if it is $1$ the first word of the list is added to the end of the word, then the list is rotated.
CTS are polynomial time Turing universal \citep{neary+woods06icalp} and have been used to provide the best known bounds on Turing machines \citep{woods+neary06focs,woods+neary07mcu,neary+woods06tcs} and cellular automata \citep{cook04:cs}.
They also provide the best bounds presented here.

\medskip

Signal machines are presented in \RefSection{sec:definitions}.
Each following section deals with a different universal model and presents ways to simulate them with SM:
Turing machines in  \RefSection{sec:turing-machines}, 
cellular automata in \RefSection{sec:ca}, and
cyclic tag systems in \RefSection{sec:cyclic-tag}.
\RefSection{sec:conclusion} gathers some concluding remarks.


\section{Definitions}
\label{sec:definitions}

In \emph{Abstract geometrical computation}, dimensionless objects are moving on the real axis.
When a collision occurs they are replaced according to rules.
This is  defined by the following machines:

\begin{Definition}
  A \emph{signal machine} is defined by $(\AGCmetaSignalSet,\AGCspeedFun,\AGCruleSet)$ where 
  \AGCmetaSignalSet (\emph{meta-signals}) is a finite set,
  \AGCspeedFun (\emph{speeds}) a mapping from \AGCmetaSignalSet to \RealSet, and
  \AGCruleSet  (\emph{collision rules}) a function from the subsets of \AGCmetaSignalSet of cardinality at least two into subsets of \AGCmetaSignalSet (all these sets are composed of meta-signals of distinct speed).
\end{Definition}

Each instance of a meta-signal is a \emph{signal}.
The mapping \AGCspeedFun assigns  \emph{speeds} to signals.
They correspond to the inverse slopes of the line segments in space-time diagrams.
A \emph{collision rule}, $\AGCrule^-{\rightarrow}\AGCrule^+$, defines what emerges ($\AGCrule^+$) from the collision of two or more signals ($\AGCrule^-$).
Since \AGCruleSet is a function, signal machines are deterministic.
The \emph{extended value set}, \AGCextendedValueSet, is the union of \AGCmetaSignalSet and \AGCruleSet plus one symbol for void, \AGCextendedValueVoid.
A \emph{configuration}, \AGCconfiguration, is a mapping from \RealSet to \AGCextendedValueSet such that the set $\{\,x\in\RealSet\,|\,\AGCconfiguration(x)\neq \AGCextendedValueVoid\,\}$ is finite.
An \emph{infinite configuration}, is a similar mapping such that the previous set has no accumulation point.

A signal corresponding to a meta-signal \AGCmetaSignal at a position $x$, \ie $c(x)=\AGCmetaSignal$, is moving uniformly with constant speed $\AGCspeedFun(\AGCmetaSignal)$.
A signal must start (resp. end) in the initial (resp. final) configuration or in a collision.
This corresponds to condition $\ref{eq:sig}$ in \RefDef{def:diagram}.
At a $\AGCrule^-{\rightarrow}\AGCrule^+$ collision, signals corresponding to the meta-signals in $\AGCrule^-$ (resp. $\AGCrule^+$) must end (resp. start) and no other signal should be present (condition $\ref{eq:col}$).

\begin{Definition}
  \label{def:diagram}
  The \emph{space-time diagram} issued from an initial configuration $\AGCconfiguration_0$ and lasting for $T$, is a mapping $c$ from $[0,T]$ to configurations (\ie from $\RealSet\times[0,T]$ to \AGCextendedValueSet) such that, $\forall (x,t)\in\RealSet\times[0,T]$~:
  \begin{enumerate}
  \item each $\{\,x\in\RealSet\,|\,\AGCconfiguration_t(x)\neq \AGCextendedValueVoid\,\}$ is finite,
  \item if $\AGCconfiguration_t(x) {=} \AGCmetaSignal$ then $\exists t_i,t_f {\in} [0,T]$ with $t_i{<}t{<}t_f$ or $0{=}t_i{\leq}t{<}t_f$ or $t_i{<}t{\leq}t_f{=}T$
    s.t.:
    \begin{itemize}
    \item $\forall t'\in(t_i,t_f)$, $\AGCconfiguration_{t'}(x+\AGCspeedFun(\AGCmetaSignal)(t'-t))=\AGCmetaSignal$\,,
    \item $t_i{=}0$ or (\,$\AGCconfiguration_{t_i}(x+\AGCspeedFun(\AGCmetaSignal)(t_i-t))=\AGCrule^-{\rightarrow}\AGCrule^+$ and $\AGCmetaSignal\in\AGCrule^+$\,),
    \item $t_f{=}T$ or (\,$\AGCconfiguration_{t_f}(x+\AGCspeedFun(\AGCmetaSignal)(t_f-t))=\AGCrule^-{\rightarrow}\AGCrule^+$ and $\AGCmetaSignal\in\AGCrule^-$\,);
    \end{itemize}
    \label{eq:sig}
  \item if $\AGCconfiguration_t(x){=}\AGCrule^-{\rightarrow}\AGCrule^+$ 
    then $\exists \varepsilon$, $0{<}\varepsilon$, $\forall t'{\in}[t{-}\varepsilon,t{+}\varepsilon]\cap[0,T]$, $\forall x'{\in}[x-\varepsilon,x+\varepsilon]$, 
    \begin{itemize}
    \item $(x',t')\neq(x,t)$ $\Rightarrow$ $c_{t'}{}(x')\in\AGCrule^-{\cup}\AGCrule^+\cup\{\AGCextendedValueVoid\}$,
    \item $\forall \AGCmetaSignal{\in}\AGCmetaSignalSet$, $c_{t'}{}(x'){=}\AGCmetaSignal$
      $\Leftrightarrow$ or 
      $\left\{
        \begin{array}{l}
          \AGCmetaSignal\in\AGCrule^- \text{ and } t'<t \text{ and } x'=x+\AGCspeedFun(\AGCmetaSignal)(t'-t)\,,
          \\
          \AGCmetaSignal\in\AGCrule^+ \text{ and } t<t' \text{ and } x'=x+\AGCspeedFun(\AGCmetaSignal)(t'-t)\,.
        \end{array}
      \right.$
    \end{itemize}
    \label{eq:col}
  \end{enumerate}
\end{Definition}

On space-time diagrams, time is increasing upward.
The traces of signals are line segments whose directions are defined by $(\AGCspeedFun(.),1)$ ($1$ is the temporal coordinate).
Collisions correspond to the extremities of these segments.
This definition can easily be extended to $T=\infty$ and to infinite initial configuration.

Although speeds may be any real and thus encode information, in the following, only a few integer values are used.
Similarly, the distance between signals may be any real but only integer positions are used.

\subsection{Time complexity measure}

As a computing device, the input is the initial configuration and the output is the final configuration.
A SM is \emph{Turing universal} if there exists encodings/representations through which it can go from the code of a Turing machine (or any equivalent model of computation) and an entry to the output (if any).
AGC provides dynamical systems with no halting feature.
Mainly two approaches exist to settle this:
\begin{itemize}
\item an observer checks that the end of the computation is reached according to some property over the configuration, \eg the presence of a  meta-signal, or
\item the system reaches a stable state, \ie no more collision is possible.
\end{itemize}
The last one is preferred since the halting is then a part of the system; although it is sometimes meaning-less: for example, when simulating a cellular automata, since there is no halting feature in CA, the same discussion arises again.

To consider polynomial time universality, the time complexity of a computation should be defined.
Space and time are continuous, by rescaling a finite computation, it can be made as short as wanted, and even worse: any infinite computation starting from a finite configuration can be automatically folded into a finite portion of the space-time diagram \citep{durand.lose09nc}.
So that a correct notion of time complexity lies elsewhere.

Collisions are considered as the discrete steps related by signals: a collision is \emph{causally before} another if a signal generated by the first one ends in the second one.
This yields a direct acyclic graph.

\begin{definition}
The \emph{time complexity} of an SM computation is the longest length of a chain in the collision causality DAG.
\end{definition}

For \emph{space complexity} one may consider the longest length of an anti-chain or the maximum number of signals present at the same time (which form a cut).

\subsection{Generating an infinite periodic signal pattern}
\label{sub:periodic-pattern}

The aim is to generate a periodic infinite sequence of signals on the side of a space-time diagram.
This is useful to generate, starting from a finite number of signals the infinite data for semi-universality (when it is periodic).
For TM, since it happens away from the head, there is no problem.
The CA case is not so simple as explained later.
In both cases, the increasing of the size allows to fall back into regular universality.

\begin{figure}[hbt]
  \centering\small\footnotesize\scriptsize
  \begin{tabular}{@{}c}
    \begin{tabular}{|c|c|}
      \hline
      \T \textbf{ Id} & \textbf{ Speed} \B \\
      \hline
      \T \SigBouncerLL & 2 \B\\
      \hline
      \T \SigBorderLOne, \SigBorderLTwo, \SigBorderLThree & 1 \B \\
      \hline
    \end{tabular} 
    \\\\
    \begin{tabular}{|c|}
      \hline
      \T \bf Collision rules\B \\
      \hline
      \T \AGCruleDef{\SigBouncerLL, \SigBorderLThree}{\SigMetaOne, \SigBorderLOne}\B \\\hline
      \T \AGCruleDef{\SigBouncerLL, \SigBorderLOne}{\SigMetaTwo, \SigBorderLTwo} \B\\\hline
      \T \AGCruleDef{\SigBouncerLL, \SigBorderLTwo}{\SigMetaThree, \SigBorderLThree}\B \\
      \hline
      \T \AGCruleDef{\SigBorderLOne, \SigMetaOne}{\SigMetaOne, \SigBorderLTwo, \SigBouncerLL} \B\\\hline
      \T \AGCruleDef{\SigBorderLTwo, \SigMetaTwo}{\SigMetaTwo, \SigBorderLThree, \SigBouncerLL}\B \\\hline
      \T \AGCruleDef{\SigBorderLThree, \SigMetaThree}{\SigMetaThree, \SigBorderLOne, \SigBouncerLL}\B \\
      \hline
    \end{tabular} 
  \end{tabular}%
  \begin{tabular}{@{}c@{}}
    \scriptsize\SetUnitlength{1.4ex}
    \begin{picture}(36,32)%
      \PSSigBouncerLLLaba(2,0)(6,2)%
      \PSSigBouncerLL(30,30)(34,32)%
      \PSSigBorderLOneLaba(0,0)(6,6)%
      \PSSigBorderLThreeLabb(4,0)(6,2)%
      \PSSigBorderLOneLabb(6,2)(14,10)%
      \PSSigBorderLOneLaba(22,22)(30,30)%
      \PSSigBorderLTwo(30,30)(32,32)%
      \PSSigBorderLOne(30,26)(36,32)%
      \PSSigMetaOneLabA(6,2)(6,32)%
      \PSSigMetaTwoLabA(14,10)(14,32)%
      \PSSigMetaThreeLabA(22,18)(22,32)%
      \PSSigMetaOneLabA(30,26)(30,32)%
      \multiput(0,0)(8,8){3}{%
        \PSSigBouncerLLLaba(6,6)(14,10)%
      }%
      \PSSigBorderLTwoLaba(6,6)(14,14)%
      \PSSigBorderLThreeLaba(14,14)(22,22)%
      \put(8,4){%
        \PSSigBorderLTwoLabb(6,6)(14,14)%
        \PSSigBorderLThreeLabb(14,14)(22,22)%
      }%
    \end{picture}    
  \end{tabular}
  \caption{Generating a periodic pattern.}
  \label{fig:periodic-pattern}
\end{figure}
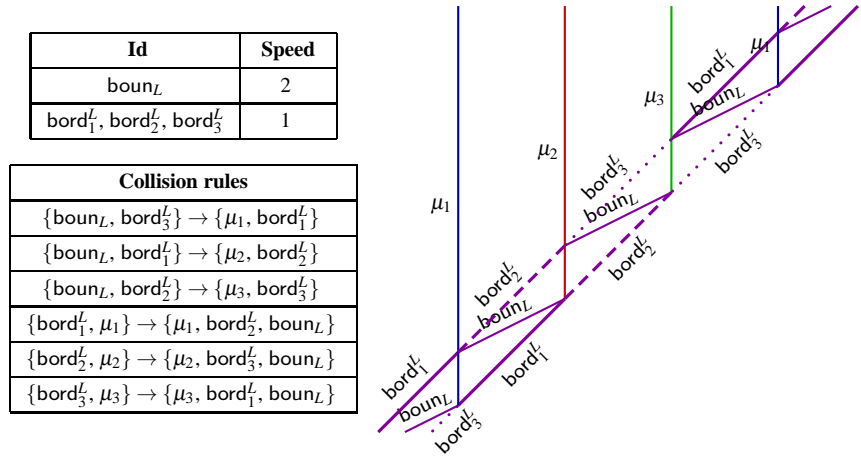

The construction works as follows: a signal bounces way and back between two signals, each time it bounces on the upper signal, a signal is emitted.
The boundary signals are used to record the location in the pattern.
The construction is simple and straightforward.
It is only presented on an example: an infinite sequence of period $3$, $(\AGCmetaSignal_1\AGCmetaSignal_2\AGCmetaSignal_3)^\omega$, where each $\AGCmetaSignal_i$ is already defined.
The added signals and rules as well as the resulting space-time diagram are displayed on \RefFig{fig:periodic-pattern}.

To save a signal, the signal is emitted at the bottom.
The number of added signals is one plus the period (the output signals are not counted).
The number of rules is twice the period.
If the output signals are to be set at unequal distances, the generator uses one plus twice the period meta-signals (one for the two borders and different way and back for the whole period).


\section{Turing machines}
\label{sec:turing-machines}

Due to the poor results generated and the lack of space, the presentation of Turing machines and the full construction from \cite{durand.lose09nc} are not given here.
The number of meta-signals in the signal machine simulating a TM is:
 1 for each tape symbol,
 2 for each state, and
 4 extra signals for enlarging the tape.
The number of collision rules is up bounded by:
 2 for each entry of the transition table of the TM, and
 plus 2 for each transition on the blank symbol plus 2 (for enlarging the table).
In the cited paper, TM halting corresponds to  the head leaving the tape. 
With a halting state,  the corresponding collision rules produce the disappearance of the state signal, at no extra cost.

Considering the curve of \cite{neary+woods06tcs,woods+neary06focs,woods+neary07mcu}, this leads to the following values:
 18 meta-signals for polynomial time universality.
The number of collision rules is bounded by 62.
The exact numbers have not been computed since fewer meta-signals is possible.

Semi-universal TM with fewer states exists.
Like the one presented in \RefSubsec{sub:periodic-pattern}, add hoc constructions to generate the extension of the tape --to achieve full universality-- would add too many states and has not been considered.
For semi-universality, using \cite{smith08cs}, the $2$-states $3$-symbol TM generates a  SM with $7$ meta-signals (the ones for enlarging are not needed) and $6$ collision rules.


\section{Cellular automata}
\label{sec:ca}

Cellular automata (CA) operate over infinite arrays of cells.
Each cell can be in finitely many states. 
(Infinite array is the only way to ensure unbounded memory.)
A CA changes the states of all the cells simultaneously according to a local function and the states of the two surrounding cells.
This is parallel, synchronous, local and uniform process.


\begin{definition}
  A cellular automaton is defined by $(\CAstateSet,\CAlocalFunction)$ where:
 \CAstateSet is a finite set of \emph{states},
 $\CAlocalFunction:\CAstateSet^3\rightarrow\CAstateSet$, is the \emph{local function}.
The \emph{global function}, $\CAglobalFunction:\CAstateSet^{\IntegerSet}\rightarrow\CAstateSet^{\IntegerSet}$, is defined by: $\forall i\in\IntegerSet, \CAglobalFunction(c)_i=\CAlocalFunction(c_{i-1},c_i,c_{i+1})$.
\end{definition}

Only CA of dimension $1$ and radius $1$ are considered here.
Higher dimensions can be covered similarly by higher dimension signal machines.
Radius $1$ means that a cell only communicates with its two closest neighbors (one on each side).
Broader radii could have been considered, but more signals are needed to convey information at greater distances (\eg $5$ meta-signals per state instead of $3$ for radius $2$).

Halting is not  provided by CA, it can be defined by reaching  a stable/periodic configuration, the apparition of a state/pattern in the configuration or on some designated cell.
In any case, more meta-signals are needed.

\begin{figure}[hbt]
  \centering
  \begin{tabular}{c}
    \large%
    \scalebox{.8}{%
      \begin{tabular}{@{\,}c@{\,}c@{\,}c@{\,}c@{\,}c@{\,}c@{\,}c@{\,}c@{\,}c@{\,}c@{\,}c@{\,}c@{\,}c@{\,}c@{\,}}
        \bf0  & 1& 1& 1& 0& 0& 1& 1& 0& \bf0    \\ 
        \bf1  &\bf0  & 1& 1& 0& 0& 0& 1&  \bf1 & \bf1\\
        & \bf1 & \bf0 & 1& 1& 1& 1&  \bf1 & \bf0 \\ 
        &  &  \bf1 &\bf0  & \it 1& \it 1& \bf0 &  \bf1 
      \end{tabular}}
  \end{tabular}
  \begin{tabular}{c}
    \includegraphics[width=.6\textwidth]{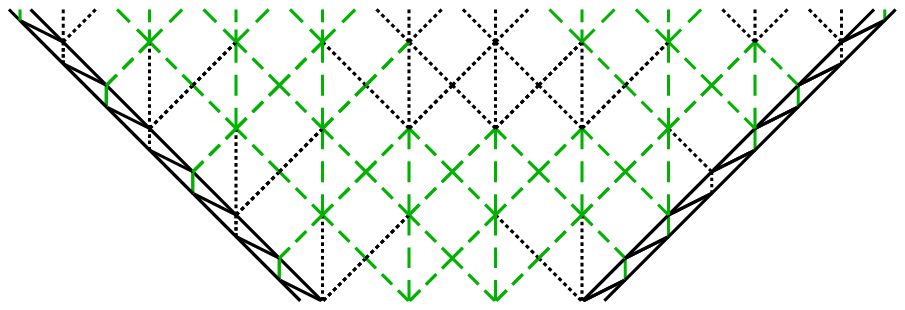}
  \end{tabular}
  \caption{Evolution and simulation of rule $110$ on $11$ framed by $^\omega(10)$ and $(011)^\omega$.}
  \label{fig:diag-110}
\end{figure}

There are two ways to manipulate finite CA-configurations:
 use some \emph{quiescent state}, \CAstateQuiscent (satisfying $\CAlocalFunction(\CAstateQuiscent,\CAstateQuiscent,\CAstateQuiscent)=\CAstateQuiscent$), for undefined cells; or
 use a periodic spatial extension on both side. 
This pattern is also (ultimately) time-periodic in the CA evolution.
In the simulation, the computation is framed by periodic signals according to oblique discrete lines in the space-time diagram of the CA as illustrated by the example on \RefFig{fig:diag-110}.

The idea is to locate the cells at integer positions and each time the local function is used, three signals are emitted, one for the cell and one for each of the closest cell on each side.
For each state, \CAstate, there are three meta-signals:
$\CAstate_L$,
$\CAstate$ and
$\CAstate_R$ or speeds $-1$, $0$ and $1$ respectively.
A transition is performed when a cell received simultaneously the values from this two neighbors (special care has to be put on locations to ensure exact meetings).
The local function is encoded in the collision rules: if $\CAlocalFunction(s,t,u)=v$ then the following rule is defined \AGCruleDef{$s_R$, $t$, $u_L$}{$v_L$, $v$, $v_R$}.
Rule $110$ is presented \RefFig{fig:def-110} as well as a generated rule.

\begin{figure}[hbt]
  \centering\scriptsize\footnotesize
  \begin{tabular}[b]{|l|@{\,}c@{\,}|@{\,}c@{\,}|@{\,}c@{\,}|@{\,}c@{\,}|@{\,}c@{\,}|@{\,}c@{\,}|@{\,}c@{\,}|@{\,}c@{\,}|}
    \hline
    \T Output  & 0 & 1 & 1 & 0  & 1 & 1 & 1 & 0 \B
    \\ \hline
    \T Input &	1\,1\,1 & 1\,1\,0 & 1\,0\,1 & 1\,0\,0 & 0\,1\,1 & 0\,1\,0 & 0\,0\,1 & 0\,0\,0  \B
    \\    \hline
  \end{tabular}
  \scriptsize\SetUnitlength{3.25em}
  \begin{tabular}[b]{c}
    \begin{picture}(3,1.7)(-2,-.8)
      \PSSigZeroRLaba(-1,-1)(0,0)
      \PSSigOneLaba(0,-1)(0,0)
      \PSSigZeroLLaba(0,0)(1,-1)
      \PSSigOneLLabb(-1,1)(0,0)
      \PSSigOneLabb(0,0)(0,1)
      \PSSigOneRLabb(0,0)(1,1)
    \end{picture}
  \end{tabular}
  \caption{Local function and the rule implementing $\CAlocalFunction(0,1,0)=1$.}
  \label{fig:def-110}
\end{figure}

Evolution and simulation on the configuration $11$, framed on the left by $^\omega(10)$ and on the right by $(011)^\omega$ is presented on \RefFig{fig:diag-110}.
As it can be seen, the way the frames are positioned (in boldface) in the evolution is not trivial and neither is the generated periodic pattern on both side of the simulation.

With the references in \cite{durand.lose09ecss} and \cite{ollinger08jac}, it is hard to go below $30$ meta-signals for full universality.
\cite{ollinger02icalp,cook04:cs} and \cite{richard+ollinger08wcsp} provide universal CA with very few states but they use periodic extensions that encode a boolean function or a cyclic tag system which leads to too many meta-signals unless a clever way to generate them is found.
Nevertheless, with rule $110$ (proved universal in \cite{cook04:cs}), a $6$ meta-signals $8$ rules semi-universal SM is generated.
This SM is not halting.
To get a halting SM from a CA, one has to know precisely what corresponds to halt and add signals and rules for the SM to halt.


\section{Cyclic tag systems simulation}
\label{sec:cyclic-tag}

Cyclic tag systems (CTS) are defined by a word and a circular list of appendants.
Both the word and the appendants are binary words.
The system is updated in the following way: the first bit of the word is removed.
If it is $1$ then the first appendant is appended at the end of the word (otherwise nothing is done).
Then, the list is rotated circularly.
The list represents the code and the word the input.
Not only are CTS able to compute \citep{cook04:cs} but also they can do it with polynomial slowdown \citep{neary+woods06icalp}.

The simulation is done with two objects:
parallel signals encoding the word, and
a cycle list: each time it cycles, depending on the signals that started it can deliver a copy of the first appendant or not.
Signals encoding the word are placed so that delivered copies automatically enlarge the word on the right.

The initial configuration is presented before the dynamics of the various elements.
The signals are defined on \RefFigure{fig:signals}.
The following naming convention is used: meta-signals with no subscript have speed $0$ and the ones with subscript $LL$, $R$ and $RR$ have speed $-2$, $1$ and $2$ respectively.

\begin{figure}[hbt]
  \centering\scriptsize\footnotesize\small
  \begin{tabular}{|c|l|}
    \hline 
    \bf Speed & \bf Meta-Signals \\
    \hline 
    -2 & \SigGoLL \\
    \hline
    0 & \SigZero, \SigOne, \SigFirst, \SigSep, \SigLast  \\
    \hline
    1 & \SigZeroR, \SigOneR, \SigFalseR, \SigTrueR  \\
    \hline
    2 & \SigZeroRR, \SigOneRR, \SigGoRR  \\
    \hline
  \end{tabular}
  \caption{List of all the meta-signals.}
  \label{fig:signals}
\end{figure}

\begin{figure}[hbt]
  \centering
  \small\SetUnitlength{1.6em}
  \newcommand{\Hei}{4}
  \begin{picture}(20,\Hei)(-2,0)
    \PSSigLastLaba(-1,0)(-1,.9)
    \PSSigGoLLLabb(-1,.9)(0.8,0)
    \PSSigGoRRLaba(-1,.9)(1,1.9)
    \PSSigLastLaba(1,1.9)(1,\Hei)
    \PSSigOneRLaba(1,1.9)(3.1,\Hei)
    \PSSigOneLabb(1,0)(1,1.9)
    \PSSigZeroLabb(2,0)(2,\Hei)
    \PSSigOneLaba(3,0)(3,\Hei)
    \PSSigOneLaba(4,0)(4,\Hei)
    \PSSigFirstLaba(5,0)(5,\Hei)
    \PSSigZeroLaba(6,0)(6,\Hei)
    \PSSigOneLaba(7,0)(7,\Hei)
    \PSSigOneLaba(8,0)(8,\Hei)
    \PSSigSepLaba(9,0)(9,\Hei)
    \PSSigOneLaba(10,0)(10,\Hei)
    \PSSigSepLaba(11,0)(11,\Hei)
    \PSSigZeroLaba(12,0)(12,\Hei)
    \PSSigOneLaba(13,0)(13,\Hei)
    \PSSigOneLaba(14,0)(14,\Hei)
    \PSSigSepLaba(15,0)(15,\Hei)
    \PSSigZeroLaba(16,0)(16,\Hei)
    \PSSigOneLaba(17,0)(17,\Hei)
    \PSSigLastLaba(18,0)(18,\Hei)
  \end{picture}
  \caption{Initial configuration and first collisions for $1011$ and list $[011,1,011,01]$.}
  \label{fig:initial}
\end{figure}
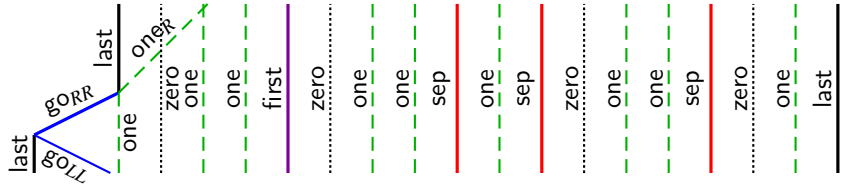

As illustrated on  \RefFigure{fig:initial}, the initial configuration is composed, left to right, of:
\SigLast, \SigGoLL that starts the dynamics, then \SigOne's and \SigZero's to encode the word, then \SigFirst to indicate the beginning of the cyclic list then alternatively  \SigOne's and \SigZero's to encode each appendant and \SigSep to separate them and finally \SigLast.
The iteration starts when the \SigGoLL signal erases \SigLast and bounces as \SigGoRR.
The later erases the first bit and send the corresponding signal, \SigZeroR or \SigOneR to cyclic list.
Once the rotation is initiated and the possible addition of signals is done, a \SigGoLL signal is sent back to the word.

The emitted \SigOneR or \SigZeroR signal crosses the \SigOne and \SigZero encoding the word doing nothing until it reaches \SigFirst where the rotation starts.
(For technical reasons, \SigFirst remains and is removed by \SigGoLL.)
If the word is empty then \SigGoRR meets \SigFirst.
Reaching the empty word is a halting condition so that, is such a case, \SigGoRR is just discarded.

\RefFigure{fig:collisions} lists all the collision rules.
All non-blank rules are collision of only two signals so that the rules can be presented in a two-dimensional array.
In blank collision rules, the output is equal to the input: the signals just cross each other unaffected.

\begin{figure}[hbt]
  \centering\scriptsize
  \begin{tabular}{@{}|@{\,\footnotesize }c@{\,}|@{\,}c@{\,}|@{\,}c@{\,}|@{\,}c@{\,}|@{\,}c@{\,}|@{\,}c@{\,}|@{}p{0cm}@{\,}c@{\,}|@{}}
    \hline 
    \T & \footnotesize \SigZero & \footnotesize  \SigOne & \footnotesize \SigFirst & \footnotesize \SigSep &\footnotesize  \SigLast &&\footnotesize  \SigGoLL
    \\ \hline 
    \T \SigGoRR &  \SigLast, \SigZeroR & \SigLast, \SigOneR & \SigFirst & & && \B
    \\ \cline{1-8}
    \T \SigOneRR &  \SigZero, \SigZeroR, \SigOneRR &  \SigOne, \SigOneR, \SigOneRR & & \SigGoLL, \SigLast, \SigFalseR & &&\B
    \\ \cline{1-8}
    \T \SigZeroRR & \SigZeroR, \SigZeroRR &  \SigOneR, \SigZeroRR & & \SigGoLL, \SigLast, \SigFalseR & &&\B
    \\ \cline{1-8}
    \T \SigTrueR &  --- & --- & & & \SigFirst, \SigFalseR && ---\B
    \\ \cline{1-8}
    \T \SigFalseR & --- & --- & \SigSep, \SigGoRR & --- & \SigFirst, \SigOneR && \SigGoLL, \SigTrueR \B
    \\ \cline{1-8}
    \T \SigOneR & --- & --- & \SigFirst, \SigTrueR, \SigOneRR & --- & --- && --- \B
    \\ \cline{1-8}
    \T \SigZeroR & --- & --- & \SigFirst, \SigFalseR, \SigZeroRR & --- & --- && --- \B
    \\ \cline{1-8}
    \T \SigGoLL & --- & --- & \SigGoLL & & \SigGoRR \B
    \\ \cline{1-6}
  \end{tabular}
  \smallskip

  \begin{tabular}{|@{\,\footnotesize }c@{\,}|@{\,}c@{\,}|@{\,}c@{\,}|@{\,}c@{\,}|@{\,}c@{\,}|}
    \hline 
    \T & \footnotesize\SigZeroR & \footnotesize\SigOneR & \footnotesize\SigFalseR & \footnotesize\SigTrueR 
    \\ \hline 
    \T \SigGoRR & & \SigLast & \SigZero, \SigGoRR & \SigOne, \SigGoRR \B
    \\ \cline{1-5}
    \T \SigOneRR &  & --- & --- & ---\B
    \\ \cline{1-5}
    \T \SigZeroRR &  & --- & --- & ---  \B
    \\ \cline{1-5}
  \end{tabular} \quad ``---'' means blank  \quad
  \begin{tabular}{|c|}
    \hline
    \T \bf Other blank rules    \B\\  \hline 
    \T \SigZeroR, \SigZero, \SigGoLL\B     \\     \hline
    \T \SigZeroR, \SigOne, \SigGoLL \B\\  \hline
    \T \SigOneR, \SigZero, \SigGoLL \B\\   \hline
    \T \SigOneR, \SigOne, \SigGoLL  \B\\   \hline
  \end{tabular}
  \caption{List of all the collision rules.}
  \label{fig:collisions}
\end{figure}

The rotation is handled in three steps:
 signals are set on movement --copies are left only if it is started by \SigOneR-- then
 the signals are moving freely to the right end of the list, and finally
 after reaching \SigLast they are positioned.
The first part is presented on \RefFigure{fig:start-rotate}.
A signal of speed $2$ (\SigZeroRR or \SigOneRR) is emitted to generate the speed $1$ versions of \SigZero's and \SigOne's encoding the first appendant.
A signal of speed $1$ is also emitted as well as another one on reaching the first \SigSep; these two signals are used to delimit the appendant during the translation.
The first speed $1$ signal is \SigFalseR for \SigZeroR and \SigTrueR for \SigOneR; so that the carried bit is preserved which is useful to finish the rotation.
The signal to start the next iteration, \SigGoLL, is emitted very quickly (it is not a problem since the next iteration cannot catch up with the rotation).
This is used to turn \SigFalseR to \SigTrueR and to provide a simple halting scheme as explained later.

\newcommand{\StartRotating}[1]{%
  \small\SetUnitlength{1.5em}%
  \begin{picture}(11,10)(1,0)%
    \PSSigFirstLabb(2,0)(2,1)
    \PSSigFirstLabb(2,1)(2,9)
    \PSSigZeroLabb(4,0)(4,2)
    \PSSigOneLaba(6,0)(6,3)
    \PSSigOneLaba(8,0)(8,4)
    \PSSigSepLaba(10,0)(10,5)
    \PSSigZeroRLaba(4,2)(11,9)
    \PSSigOneRLaba(6,3)(11,8)
    \PSSigOneRLaba(8,4)(11,7)
    \PSSigLastLabb(10,5)(10,9)
    \PSSigFirstLaba(10,9)(10,10)
    \PSSigGoLLLaba(1,9.5)(10,5)
    \PSSigFalseRLabb(10,5)(11,6)
    \PSSigFalseRLabb(10,9)(11,10)
    #1
  \end{picture}%
}

\begin{figure}[hbt]
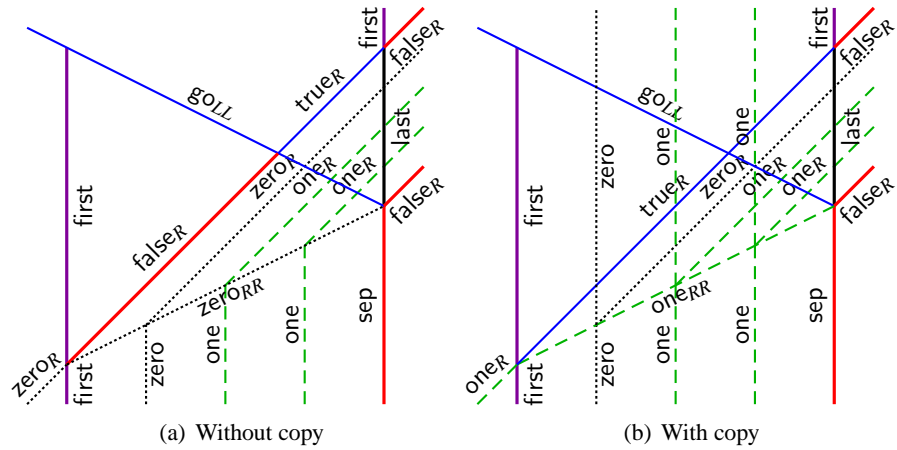

  \centering
  \subfigure[Without copy]{\StartRotating{%
    \PSSigZeroRLaba(1,0)(2,1)%
    \PSSigFalseRLaba(2,1)(7.33,6.33)%
    \PSSigTrueRLaba(7.33,6.33)(10,9)%
    \PSSigZeroRRLabb(2,1)(10,5)%
  }}
  \subfigure[With copy]{\StartRotating{%
    \PSSigOneRLaba(1,0)(2,1)%
    \PSSigOneRRLabb(2,1)(10,5)%
    \PSSigTrueRLaba(2,1)(10,9)%
    \PSSigZeroLabb(4,2)(4,10)
    \PSSigOneLaba(6,3)(6,10)
    \PSSigOneLaba(8,4)(8,10)
  }}
  \caption{Starting the rotation and restarting the dynamics.}
  \label{fig:start-rotate}
\end{figure}

From then, the rest of the translation is same in both cases.
The middle part is straightforward: the parallel \SigFalseR, \SigZeroR and \SigOneR are crossing \SigSep, \SigZero and \SigOne until \SigLast is met.

\begin{figure}[hbt]
  \centering
  \small\footnotesize\SetUnitlength{1.6em}%
  \begin{tabular}{@{}c@{}}
    \begin{picture}(12.5,9.5)(0,1)%
      \PSSigLastLabb(2,1)(2,2)
      \PSSigFirstLaba(2,2)(2,6)
      \PSSigSepLaba(2,6)(2,10.5)
      \PSSigZeroLaba(4,7)(4,10.5)
      \PSSigOneLaba(6,8)(6,10.5)
      \PSSigOneLaba(8,9)(8,10.5)
      \PSSigLastLabb(10,10)(10,10.5)
      \PSSigFalseRLaba(0,4)(2,6)
      \PSSigZeroRLaba(0,3)(2,5)
      \PSSigOneRLaba(0,2)(2,4)
      \PSSigOneRLaba(0,1)(2,3)
      \PSSigFalseRLaba(1,1)(2,2)
      \PSSigGoRRLaba(2,6)(10,10)
      \PSSigFalseRLabb(2,5)(4,7)
      \PSSigTrueRLabb(2,4)(6,8)
      \PSSigTrueRLabb(2,3)(8,9)
      \PSSigOneRLabb(2,2)(10,10)
      \PSSigZeroRR(2,5)(8,8)
      \PSSigZeroRRLabb(8,8)(12,10)
      \PSSigOneRR(2,4)(6,6)
      \PSSigOneRRLabb(6,6)(12,9)
      \PSSigOneRR(2,3)(4,4)
      \PSSigOneRRLabb(4,4)(12,8)
    \end{picture}%
  \end{tabular}
  \caption{Ending the rotation.}
  \label{fig:rotate-end}
\end{figure}
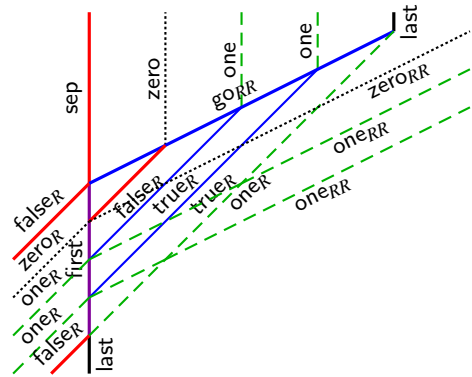

The last part of the rotation could have been the symmetric of the first part; in some ways, it is as presented on \RefFigure{fig:rotate-end}. 
To save meta-signals, \SigFalseR have been used both to mark the beginning and the end of the appendant.
But they have different meanings, so the first one changes \SigLast to \SigFirst and the second to \SigSep and \SigGoRR.
The problem is that \SigFirst interacts with \SigZeroR and \SigOneR: \SigZeroR (resp. \SigOneR) is changed to \SigFalseR (resp. \SigTrueR) and \SigOneRR (resp. \SigZeroRR).
This generates the lattice in the triangle (the signals do not interact inside it).
The bits of the appendant are now encoded with \SigFalseR's and \SigTrueR's and follow the paths \SigZeroR's and \SigOneR's would have.
It remains to \SigGoRR to turn them back into \SigZero's and \SigOne's and to turn the final \SigOneR to \SigLast.
\SigGoRR, \SigZeroRR and \SigOneRR are parallel, all the translating signals are parallel; so that the appendant is recreated with exactly the same distances.  

\begin{figure}[hbt]
  \centering
  \begin{tabular}{@{}c@{}}
    \subfigure[One CTS iteration. \label{fig:sim-tag:one-cycle}]{\includegraphics[width=.5\textwidth]{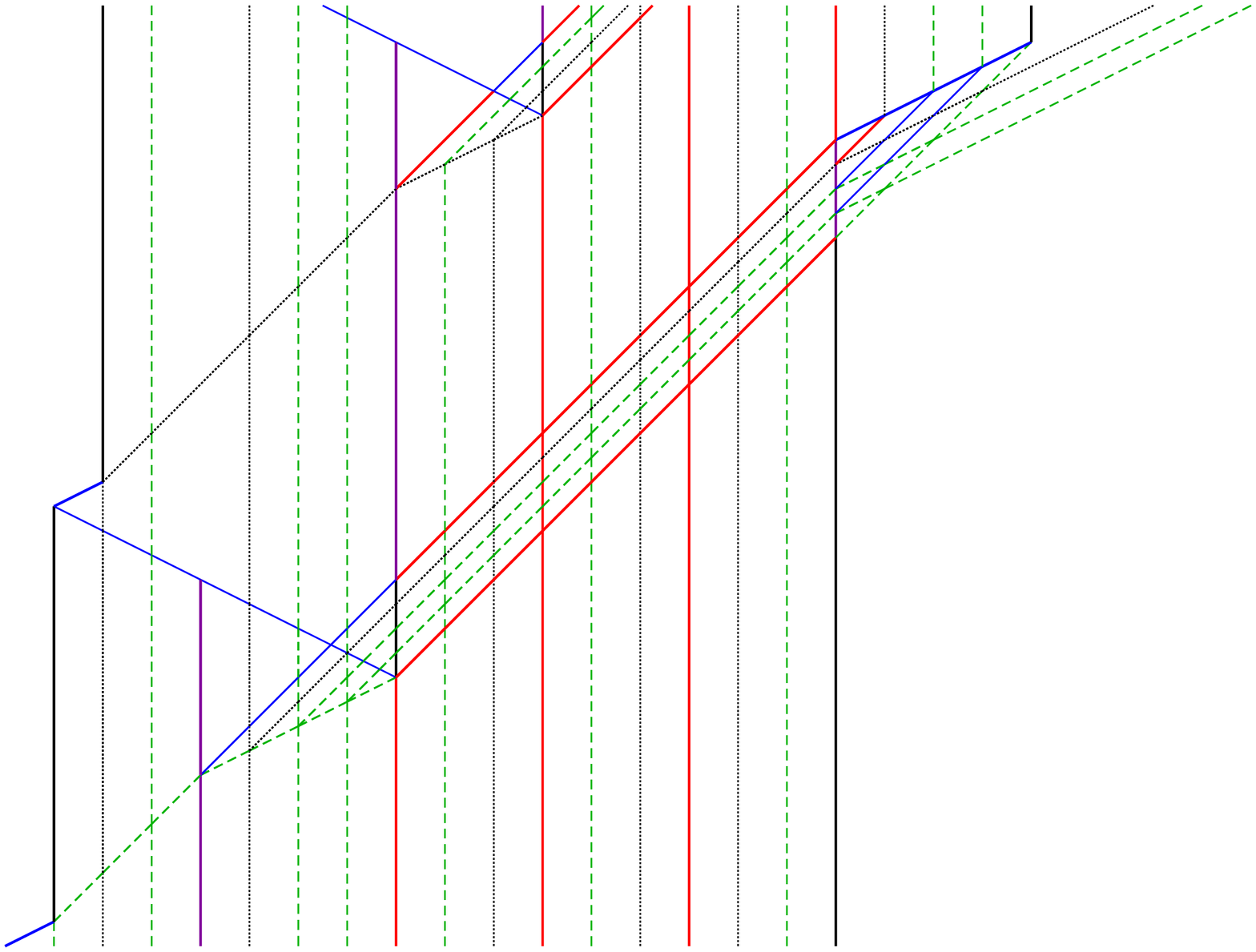}}
    \\
    \subfigure[Halt by empty word. \label{fig:sim-tag:halt-empty}]{\includegraphics[width=.34\textwidth]{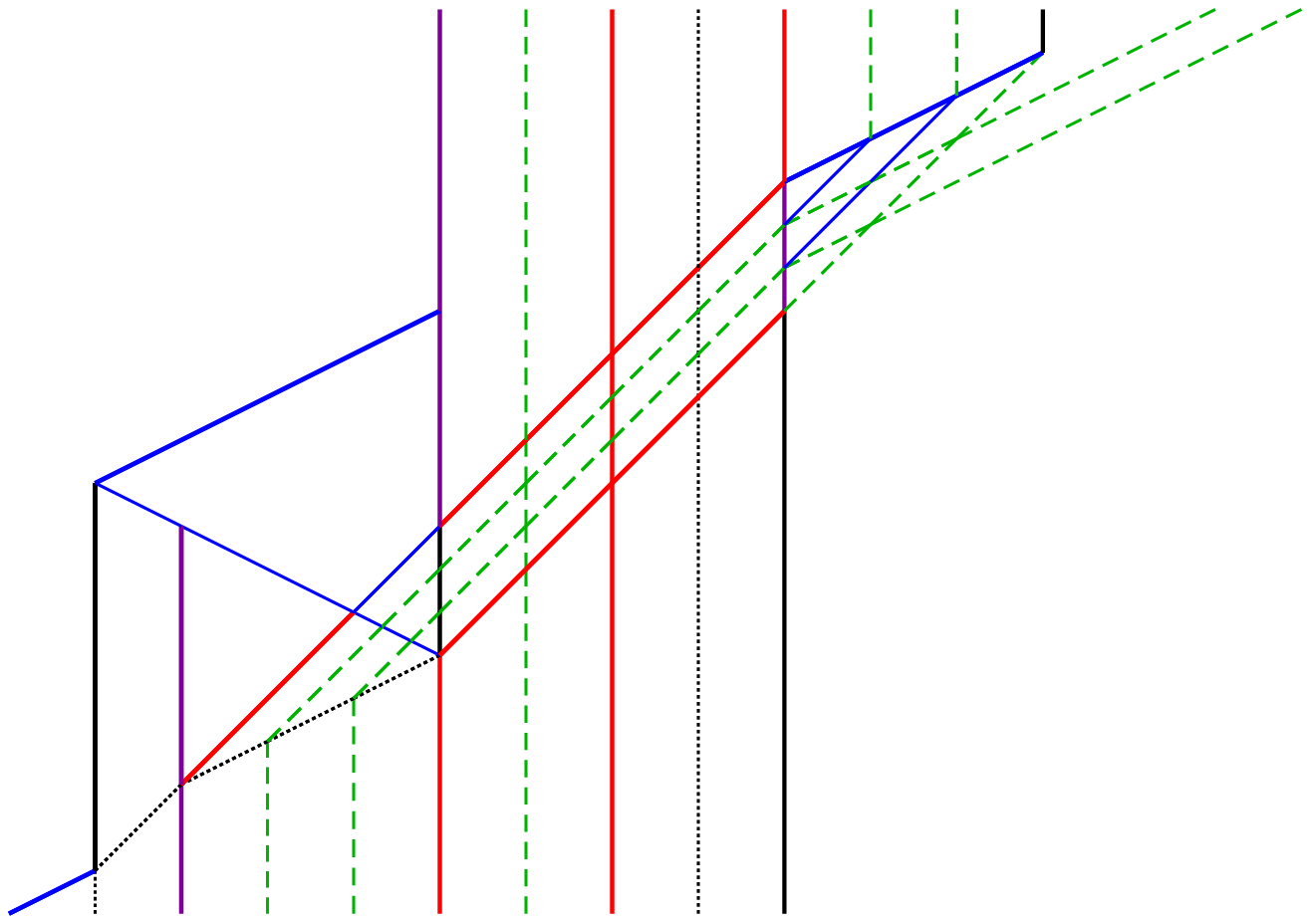}}
    \\
    \subfigure[Use of halt appendant. \label{fig:sim-tag;halt-appendant}]{\includegraphics[width=.5\textwidth]{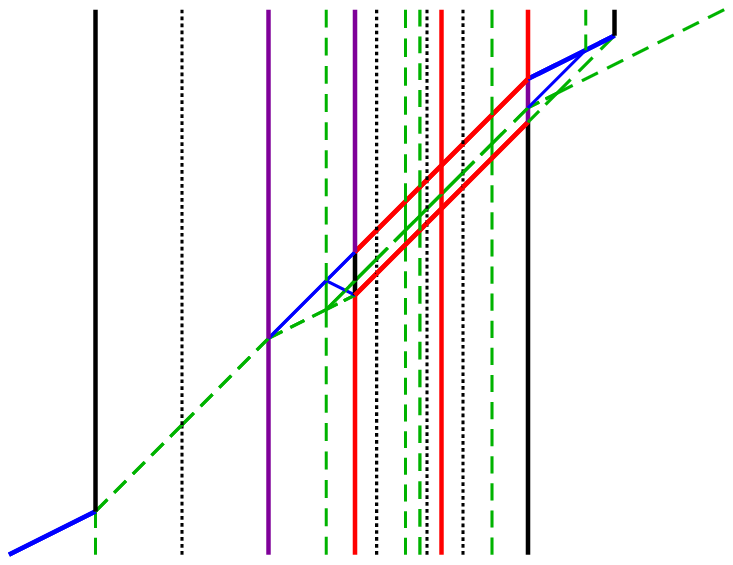}}
  \end{tabular}
  \begin{tabular}{cc}
    \subfigure[A full simulation (clean \& halt). \label{fig:sim-tag:whole}]{\ \ \includegraphics[height=.7\textheight]{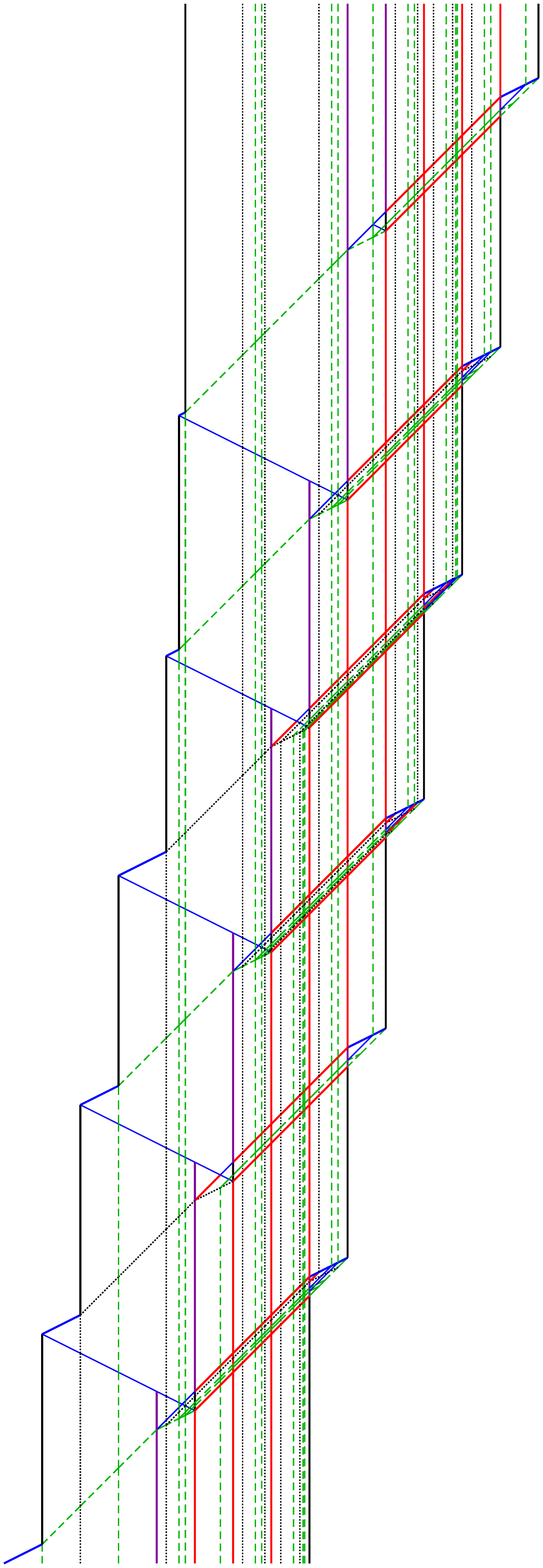}\ }
  \end{tabular}
  \caption{Simulation of a cyclic tag system.}
  \label{fig:sim-tag}
\end{figure}

\RefFigure{fig:sim-tag:one-cycle} shows one full iteration of a CTS including a whole rotation.
\RefFigure{fig:sim-tag:halt-empty} shows the halting by reaching an empty word.
\RefFigure{fig:sim-tag;halt-appendant} shows the effect of the halting appendant as explained below.
\RefFigure{fig:sim-tag:whole} shows one entire  simulation with a halting appendant and cleaning added: blank rules have been modified in order to destroy the garbage signals escaping on the right that nevertheless  would never interact with the rest of the configuration nor provoke any collisions since they are parallel.

\subsection{Adding halt at the cost of one rule}

Three halting conditions exist: on empty word, on cycling or on special halting appendant.
The first does not yield any result but nevertheless has to be implemented and already is.
The second needs an extra layer of construction to detect cycling and is not considered.
In the third case, when the halt appendant is activated, \SigGoLL has to meet \SigTrueR.
The position of this collision is determined by the distance between \SigFirst and the next \SigSep.
It appends exactly at $2/3$ of the distance.
If a \SigOne signal is set exactly at this position, it gets into the collision.
The extra rule
\[
\AGCruleDef{\SigTrueR,\SigOne,\SigGoLL}{\SigTrueR}
\]
 is used to destroy both \SigOne and \SigGoLL.
Then the rotation finished and no more collision is possible.
The result of the computation is the sequence left of \SigFirst.

The rotating process ensures that the distances between the signals remain constant, so that a \SigOne at $2/3$ remains there (and a \SigOne not at $2/3$ cannot get to  $2/3$).
It should also be ensured that, in the initial position, there is no  \SigOne  at $2/3$ not standing for halt which is straightforward to reach by, \eg, using locations $1/2$, $3/4$, $7/8$\dots


\section{Conclusion}
\label{sec:conclusion}

\begin{theorem}
  There is a universal halting signal machine with  $13$ meta-signals and $21$ non-blank rules.
  There is a non-halting semi-universal signal machine with  $6$ meta-signals and $8$ non-blank rules.
\end{theorem}

The first construction uses a collision with $3$ signals.
A $3$-signal collision needs perfect synchrony while $2$-signal collisions are more robust to small perturbations.
If only $2$ signals collisions are allowed, then a halting meta-signals can be added and processed in the circular list as the value $1$ (except that it does not generate other signals in the lattice at the end of the rotation).
This leads to a signal machine with $15$  meta-signals (the same ones plus \SigHalt and \SigHaltR) and $24$ non blank rules.

The race for small universal devices also runs for restrictions, \eg to reversible machines.
Universality has also been provided for reversible and conservative SM \cite{durand.lose06cie06}.
We believe that the CTS simulation could be turned reversible with less than two extra meta-signals.

One key feature of AGC is that the continuous space and time can be used to produce accumulations and on top of it to do black hole computations (an accumulation containing a whole, potentially infinite, Turing computation).
One may wonder about the minimal number of meta-signals for an accumulation and for the black hole effect.
Let us note that $4$ signals is enough to make an accumulation and $3$ might to be enough, whereas $2$ is not.
It also seems that the black hole effect can be added to the cyclic tag system simulation with a few extra meta-signals.


\small

\bibliographystyle{eptcs}
\makeatletter \@ifundefined{mathbb}{\long\def\mathbb{\mathsf}}
  \makeatother\providecommand{\href}[2]{#2}



\end{document}